\documentclass[sigconf]{acmart}

 \usepackage{balance}
\usepackage{tabularx}
\usepackage{booktabs}
\usepackage{multirow}
\usepackage{adjustbox}
\usepackage{varwidth}
\usepackage{algorithm}
\usepackage{graphicx}
\usepackage{xcolor}
\usepackage{hyperref}
\usepackage{caption}
\usepackage{enumitem}
\usepackage{mdframed}
\usepackage{xcolor}
\usepackage{lipsum}
\usepackage{tcolorbox} 
\usepackage{pifont}
\usepackage{diagbox}
\usepackage[flushleft]{threeparttable}
\usepackage{tikz}
\usepackage{ulem}
\usepackage{flushend}
\usepackage{xcolor}
\usepackage{tablefootnote}
\usepackage{makecell}
\usepackage{caption}
\usepackage{subcaption}
\usepackage[utf8]{inputenc}
\usepackage{microtype}
\usepackage{amsmath} 
\usepackage{utfsym}
\usepackage{longtable}
\usepackage{algpseudocode}
\usepackage{nameref}
\usepackage{listings}
\usepackage{booktabs}
\usepackage{pifont}        

\lstdefinelanguage{MyLang}{
    keywords={@Persona,@Instructions,@InputVariable,@Command,@OutputVariable,@Format,@Examples,@Terminology,@Term},
    sensitive=true,
    comment=[l]{//},
    morecomment=[s]{/*}{*/},
    morestring=[b]",
    keywordstyle=\bfseries     
}

\newcommand{\find}[1]{%
  \begin{tcolorbox}[
      colback=gray!10,   
      colframe=gray!70,  
      leftrule=1mm,
      toprule=0mm,
      bottomrule=0mm,
      left=1pt,
      right=2pt,
      top=2pt,
      bottom=2pt
  ]
  \textit{#1}
  \end{tcolorbox}%
}

\newcommand{\hiddenref}[1]{\hyperref[#1]{}}
\newcommand{\tool}[0]{SemFuzz}%

\begin{document}

\title{SemFuzz: A Semantics-Aware Fuzzing Framework for Network Protocol Implementations}

\author{Yanbang Sun}
\affiliation{
  \institution{Tianjin University}
  \city{Tianjin} 
  \country{China}
}
\email{ybsun@tju.edu.cn}

\author{Quan Luo}
\affiliation{
  \institution{QI-ANXIN Codesafe Team}
  \city{Beijing} 
  \country{China}
}
\email{a4651386@163.com}

\author{Yuelin Wang}
\affiliation{
  \institution{Tianjin University}
  \city{Tianjin} 
  \country{China}
}
\email{violin0613@tju.edu.cn}

\author{Qian Chen}
\affiliation{
  \institution{QI-ANXIN Codesafe Team}
  \city{Beijing} 
  \country{China}
}
\email{cq674350529@gmail.com}

\author{Benjin Liu}
\affiliation{
  \institution{QI-ANXIN Codesafe Team}
  \city{Beijing}
  \country{China}
}
\email{liubenjin@qianxin.com}

\author{Ruiqi Chen}
\affiliation{
  \institution{QI-ANXIN Codesafe Team}
  \city{Beijing}
  \country{China}
}
\email{chenruiqi@qianxin.com}

\author{Qing Huang}
\affiliation{
  \institution{Jiangxi Normal University}
  \city{Nanchang} 
  \country{China}
}
\email{qh@whu.edu.cn}

\author{Xiaohong Li\textsuperscript{*}}
\thanks{\textsuperscript{*}co-corresponding author.}
\affiliation{
  \institution{Tianjin University}
  \city{Tianjin} 
  \country{China}
}
\email{xiaohongli@tju.edu.cn}

\author{Junjie Wang\textsuperscript{*}}
\affiliation{
  \institution{Tianjin University}
  \city{Tianjin} 
  \country{China}
}
\email{junjie.wang@tju.edu.cn}

\renewcommand{\shortauthors}{Yanbang Sun et al.}

\begin{abstract}
Network protocols are the foundation of modern communication, yet their implementations often contain semantic vulnerabilities stemming from inadequate understanding of specification semantics.
Existing gray-box and black-box testing approaches lack semantic modeling of protocols, making it difficult to precisely express testing intent and cover boundary conditions.
Moreover, they typically rely on coarse-grained oracles such as crashes, which are inadequate for identifying deep semantic vulnerabilities.
To address these limitations, we present a semantics-aware fuzzing framework, SemFuzz.
The framework leverages large language models to extract structured semantic rules from RFC documents and generates test cases that intentionally violate these rules to encode specific testing intents. It then detects deep semantic vulnerabilities by comparing the observed responses with the expected ones.
Evaluation on seven widely deployed protocol implementations shows that SemFuzz identified sixteen potential vulnerabilities, ten of which have been confirmed. Among the confirmed vulnerabilities, five were previously unknown and four have been assigned CVEs.
These results demonstrate the effectiveness of SemFuzz in detecting semantic vulnerabilities.
\end{abstract}

\maketitle

\section{Introduction}

Network protocols are the cornerstone of modern communication systems, supporting data exchange across various public services.
However, flaws in their implementations often evolve into severe security threats.
This issue is particularly critical for closed-source protocol stacks, which are widely deployed in sensitive domains such as government agencies, industrial control systems, and healthcare infrastructures.
Once a flaw is introduced, its impact can propagate on a massive scale.
For instance, the Windows IPv6 protocol stack (\texttt{tcpip.sys}) once exposed  the Bad Neighbor vulnerability~\cite{CVE-2020-16898}, exposing nearly one billion devices to remote denial-of-service (DoS) attacks.
Recent studies~\cite{oday, zhang2024survey} further show that such vulnerabilities are increasing in number, underscoring the urgent need for more effective detection techniques.

\begin{figure}[t]
\centering
\includegraphics[width=0.96\linewidth]{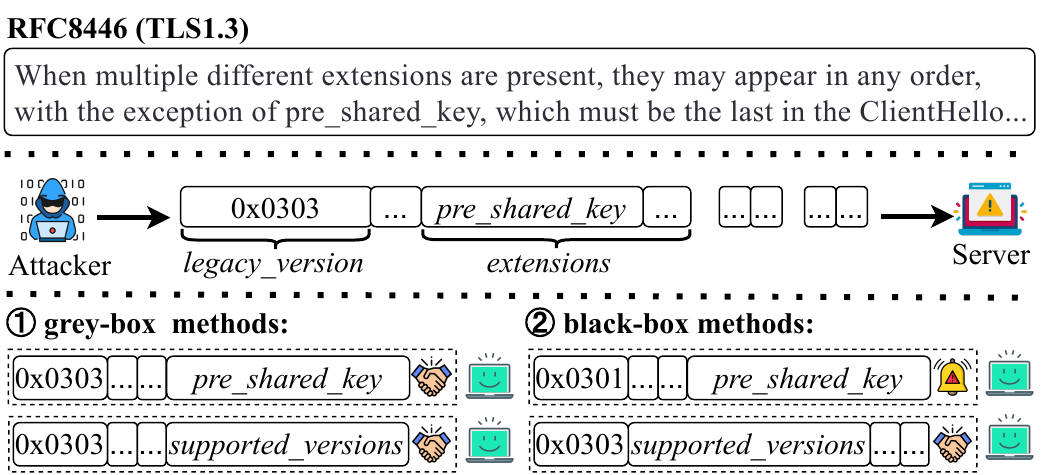}
\captionsetup{justification=centering}
\vspace{-2mm}
\caption{Limitation of existing methods}
\label{fig: killing example}
\vspace{-2mm}
\end{figure}

In general, existing methods fall into two categories: grey-box~\cite{TLSFuzzer, CHATAFLPAPER, TCPFUZZ, AFLNET, BLEEM, 10.1145} and black-box testing~\cite{Peach, HdiffPaper, chen2018rfc, Boofuzz}.
Despite the significant success of these methods, they still suffer from two inherent limitations when detecting semantic vulnerabilities in protocol implementations, especially in closed-source systems.
First, they lack semantic awareness, making it difficult to generate test cases that effectively cover boundary scenarios.
Second, they rely on coarse-grained oracles, which leads to the overlook of many semantic vulnerabilities that do not immediately trigger crashes.

Fig.~\ref{fig: killing example} illustrates a typical example: according to Section 4.2 of Request for Comments (RFC) 8446~\cite{TLS_PROTOCOL}, when the \textit{pre\_shared\_key} extension is present in a \textit{ClientHello} message, it must appear as the last extension in the extension list.
However, the TLS implementation on the Windows platform (\texttt{schannel.dll}) does not strictly adhere to this requirement.
An attacker can craft a message with this extension placed in a non-terminal position and still complete the handshake.
After a long period of specific interactions, this behavior may cause the program to crash, leading to a DoS attack.

Existing methods struggle to detect such semantic vulnerabilities.
specificity, gray-box approaches (Fig.~\ref{fig: killing example}-\ding{172}) rely on runtime signals such as coverage to drive mutations, but stable binary instrumentation is often infeasible for closed-source protocol stacks~\cite{BLEEM};
once feedback is lost, gray-box tools degrade into random or grammar-level mutators.
Moreover, due to the lack of semantic knowledge, they struggle to construct boundary test cases such as reordering of extension fields, causing vulnerabilities to be missed.
Black-box methods (Fig.~\ref{fig: killing example}-\ding{173}) do not depend on instrumentation, but typically only model syntactic structure and likewise lack an understanding of semantics.
They can mutate field values (e.g., \textit{legacy\_version}) to trigger alerts, but still fail to detect this vulnerability.
Even with manually crafted templates generating semantically violating inputs (the second packet in \ding{173}), the coarse-grained oracles (e.g., crash) often fail to identify semantic deviations that do not manifest immediately, leading to missed vulnerabilities.

In this paper, we propose \tool{}, a semantics-aware black-box fuzzing framework for protocol implementations to detect deep semantic vulnerabilities.
The core idea is to leverage large language models (LLMs) to extract semantic knowledge from RFC documentation and employ it to guide a closed-loop testing process consisting of semantic modeling, intent-driven mutation, and response verification.
Specifically, to compensate for the lack of semantic awareness, \tool{} introduces semantic modeling, leveraging an LLM to parse RFC documents and extract structured semantic rules.
Importantly, each rule specifies both construction constraints, which describe valid message formats, and processing expectations, which define the expected response that these messages should trigger.
It converts unstructured natural language descriptions into precise, machine-readable rules, thus establishing a solid semantic foundation for subsequent testing.

Based on the extracted semantic rules, \tool{} adopts an intent-driven mutation strategy to generate targeted test cases.
Specifically, it applies atomic operations (such as adding, deleting, or updating fields) to seed messages captured from real traffic according to the construction constraints of a given semantic rule, thereby generating test inputs that deliberately violate the semantic constraints.
This ensures that every test case carries a clear testing intent, aiming to evaluate the behavior of protocol implementations under specific boundary conditions.
Finally, to address the oracle problem, we introduce a precise semantic oracle to detect semantic vulnerabilities.
For each test case that is designed to violate a specific construction constraint, \tool{} extracts its corresponding processing expectation from the semantic rule and treats it as the expected response for that test.
Our oracle is then defined as follows: if the target’s actual response deviates from this expected response, a potential vulnerability is identified.
This specification-based response comparison mechanism enables \tool{} to go beyond traditional crash or memory error detection, thereby systematically uncovering a wide range of critical semantic vulnerabilities.

In the systematic evaluation across seven protocol implementations, \tool{} demonstrated exceptional vulnerability detection capability, identifying 16 potential vulnerabilities, of which 10 were real vulnerability, achieving a detection accuracy of 62.5\%.
Through responsible disclosure, 10 vulnerabilities have been confirmed, including 5 previously unknown ones (with 4 CVEs).
\tool{} significantly outperforms existing approaches, with the best baseline detecting only 5 vulnerabilities.
Ablation studies demonstrated the effectiveness of the framework’s design: the semantic rule constructor improved semantic modeling accuracy by 5.3\% and contributed to the discovery of 2 vulnerabilities, while the test case generator significantly enhanced testing effectiveness, increasing test accuracy by 142\% and uncovering 8 additional real vulnerabilities.
Multi-model experiments also show that \tool{}’s performance stems from its reasonable design rather than any specific LLM.

This paper makes the following contributions: 
\begin{itemize}[leftmargin=*]
\item 
\textbf{A semantic-aware fuzzing paradigm} is proposed, which leverages LLMs to extract protocol semantics from RFCs, transforming unstructured semantic knowledge into executable testing intents to address existing limitations.
\item
\textbf{A closed-loop workflow} integrating semantic modeling, intent mutation, and response validation is designed, where semantic rules guide the generation of violating test cases and the construction of precise semantic oracles, enabling efficient detection of deep semantic vulnerabilities.
\item 
\textbf{A systematic evaluation} across seven widely deployed protocol implementations demonstrates the superior effectiveness of \tool{}, which discovered 10 real vulnerabilities, including 5 previously unknown ones (4 assigned CVEs).
\end{itemize}
\section{Background and Motivation}

\subsection{Preliminary}
RFCs~\cite{HTTP_PROTOCOL, TLS_PROTOCOL, IP_PROTOCOL, DNS_PROTOCOL} specify message structures, field constraints, and error-handling procedures, serving as the basis for protocol implementations.
However, these specifications are written in natural language for human understanding, making it difficult for existing testing methods to leverage their semantic information directly.
To facilitate subsequent semantic modeling, we formalize the RFC specification as a set of requirements:
\[
\mathcal{D} = \{ \mathcal{R}_1, \ldots, \mathcal{R}_n \}
\]
where each $\mathcal{R}_i = (p, m, c)$ represents the protocol name, message type, and specific content.

\begin{figure}[t]
\centering
\includegraphics[width=0.96\linewidth]{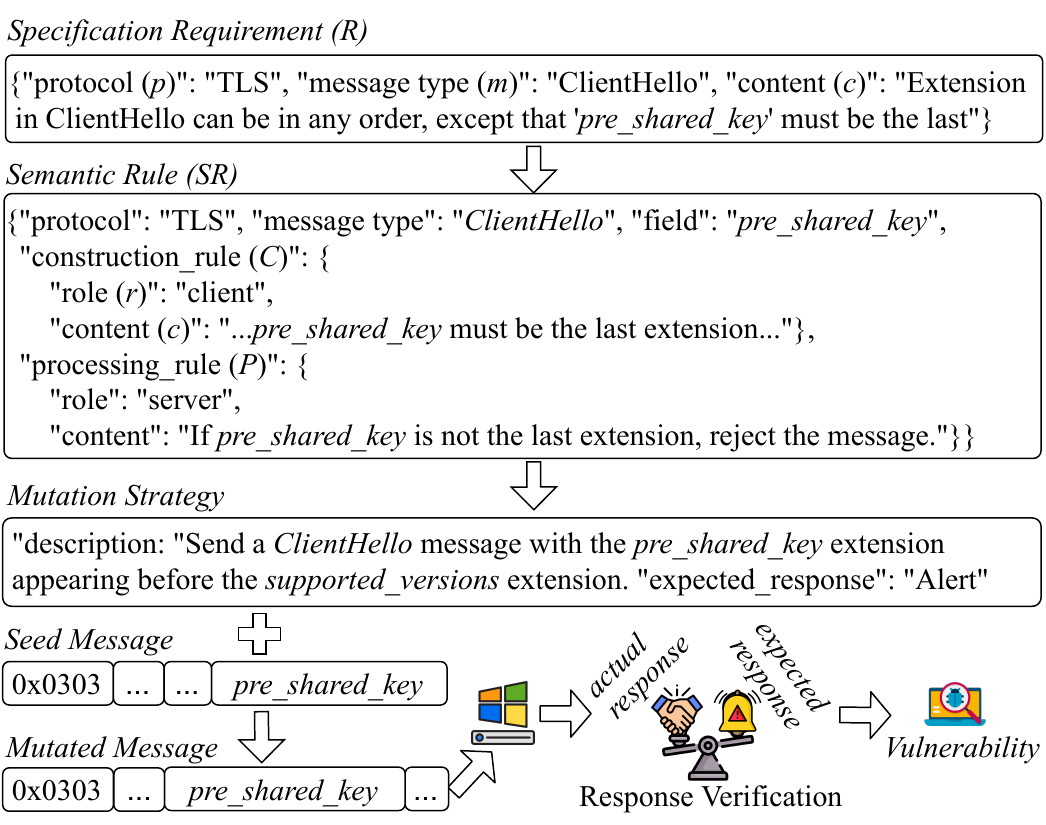}
\captionsetup{justification=centering}
\vspace{-2mm}
\caption{Motivation example}
\label{fig: motivation_example}
\vspace{-2mm}
\end{figure}

\subsection{Motivation Example} \label{sec: motivation}
The key insight of our approach is to leverage the semantic knowledge in RFCs to guide the generation of targeted test cases and to identify potential semantic vulnerabilities by comparing the expected and actual responses.
Formally, a specification requirement $\mathcal{R}_i$ can be converted to a semantic rule $\mathcal{SR}$, which written as
\[ \mathcal{SR}_i = (p, m, f, \mathcal{C}, \mathcal{P}) \] 
where $p$, $m$, and $f$ are the protocol, message, and field names; $\mathcal{C}$ is the construction constraint and $\mathcal{P}$ is the processing expectation.
The construction constraint is represented as $\mathcal{C}=(r, c)$, where $r$ denotes the role (e.g., client) and $c$ represents the specific content.
$\mathcal{P}$ follows the same format as $\mathcal{C}$.

The motivation example of our approach is illustrated in Fig.~\ref{fig: motivation example}.
We first extract specification requirements $\mathcal{R}$ from RFC documents and convert them into semantic rules $\mathcal{SR}$ which explicitly define both construction rules (i.e., what the client should send) and processing rules (i.e., how the server should respond).
Based on these rules, we generate mutation strategies that intentionally violate specific construction constraints, e.g., placing the \textit{pre\_shared\_key} extension before the \textit{supported\_versions} extension.
Each strategy specifies the expected server response (e.g., Alert).
By comparing the server's actual response with the expected result, we successfully detected a semantic vulnerability.
\section{METHODOLOGY}

\subsection{Overview}
\begin{figure}[t]
\centering
\includegraphics[width=1.0\linewidth]{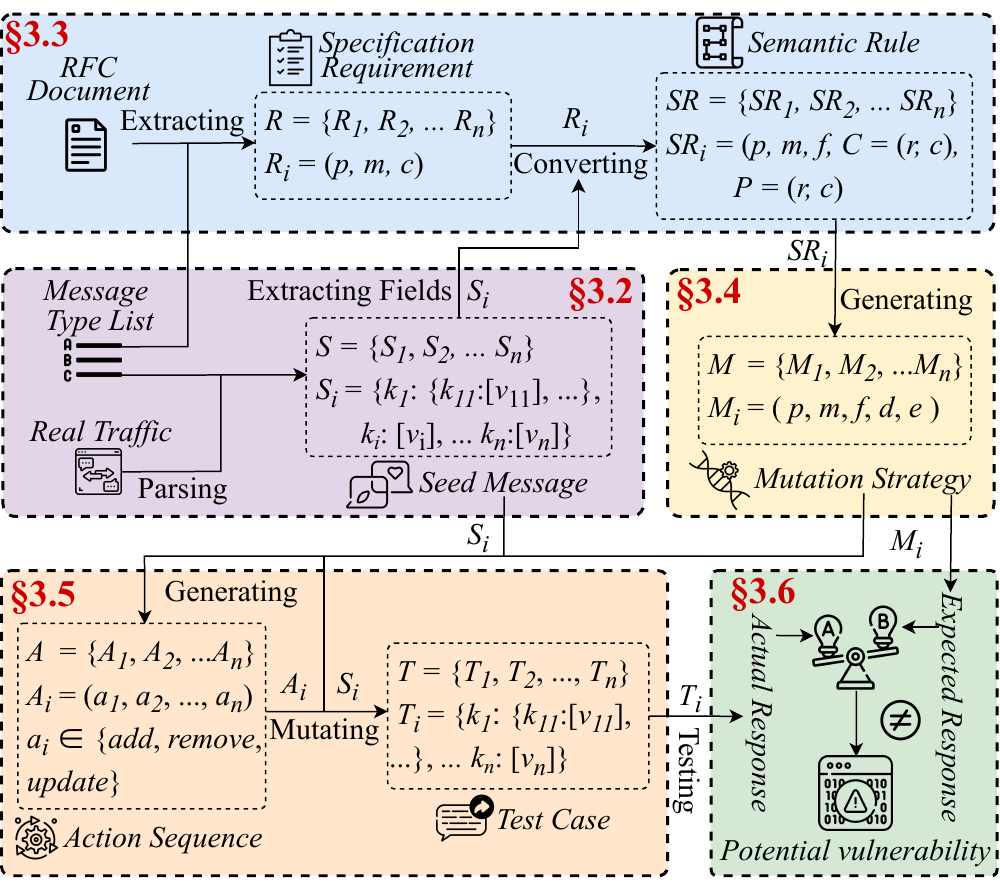}
\vspace{-4mm}
\captionsetup{justification=centering}
\caption{Overall architecture of \tool{}}
\label{fig: approach}
\vspace{-2mm}
\end{figure}

To enable semantic vulnerability detection, we design a five-stage black-box fuzzing framework, \tool{} (Fig.~\ref{fig: approach}): traffic collector (\S\ref{sec: Traffic Collector}), structured rule constructor (\S\ref{sec: Rule Constructor}), mutation-strategy generator (\S\ref{sec: Strategy Generator}), test-case generator (\S\ref{sec: Test Case Generator}), and response verifier (\S\ref{sec: Response Verifier}).

Specifically, as shown in Fig.~\ref{fig: approach}, traffic collector begins with the collection of real-world traffic based on the given message type list ($\mathcal{L}$), from which we construct a set $\mathcal{S}$ of seed messages:
\[\mathcal{S} = \{ \mathcal{S}_1,\; \mathcal{S}_2,\; \dots,\; \mathcal{S}_n \}\]
Each seed $\mathcal{S}_i$ is represented as a mapping from field keys to values:
$\mathcal{S}_i= (k_{1}:[v_{1}],\; \dots,\; k_{n}:[v_{n}])$, 
where $k_i$ denotes a message field and $v_{i}$ are observed values for that field.
Meanwhile, the structured rule constructor uses a LLM to parse the RFC text, extract requirements $\mathcal{R}$, and further convert them into semantic rule $\mathcal{SR}$.

For a selected $\mathcal{SR}_i$, the mutation strategy generator generates one or more candidate mutation strategies:
\[ \mathcal{M} = \{ \mathcal{M}_1,\; \mathcal{M}_2,\; \dots,\; \mathcal{M}_n \}\]
where each $\mathcal{M}_i = (p,\; m,\; f,\; d,\; e)$ encodes a specific strategy to violate the construction rule $\mathcal{C}$ (ref. \S\ref{sec: motivation}).
Here, $d$ is a textual description of the mutation strategy, derived from the violation of \(\mathcal{C}\);
$e$ denotes the expected behavior upon violation, e.g., Alert.

Based on the mutation strategies $\mathcal{M}$ and the seed message $\mathcal{S}$, \tool{} then generates a set of action sequences:
\[ \mathcal{A} = \{ \mathcal{A}_1,\; \mathcal{A}_2,\; \dots,\; \mathcal{A}_n \}\]
The action sequence $\mathcal{A}_i$ consists of a series of atomic action, such as \texttt{add}, \texttt{remove}, or \texttt{update}.
Applying $\mathcal{A}_i$ to the corresponding seed message ($\mathcal{S}_i$) then yields a set of test cases:
\[
\mathcal{T} = \{ \mathcal{T}_1,\; \mathcal{T}_2,\; \dots,\; \mathcal{T}_n\}
\]
Each test case $\mathcal{T}_i$ maintains syntactic validity while intentionally violating the construction rule $\mathcal{C}$.
Finally, the set of test cases $\mathcal{T}$ is sent to the target protocol implementation.
For each $\mathcal{T}_i$, we record its actual response ($r$) and compare it with the expected response ($e$) of the mutation strategy $\mathcal{M}_i$.
If the two are inconsistent, the implementation is considered to have a potential vulnerability.

\subsection{Traffic Collector}\label{sec: Traffic Collector}
To ensure the realism and validity of test inputs, we design a traffic collector to automatically extract real-world seed messages $\mathcal{S}$ from network environments.
The traffic collection process begins with specifying the message type and the IP address of the target application.
We then trigger interactions between the client and the server by executing predefined scripts, thereby stimulating message exchanges.
These interactions are captured in real time using Wireshark~\cite{wireshark_about}, and parsed into structured records via its command-line interface, tshark.
Each parsed message is treated as a seed $\mathcal{S}_i$ and added to the seed set $\mathcal{S}$ for subsequent mutation.

\begin{algorithm}[t]
\raggedright
\caption{Structured Rule Construction}
\label{alg:sr-constructor}
\textbf{Input:} RFC document $\mathcal{D}$, message type list $\mathcal{L}$, seed messages $\mathcal{S}$ \\
\textbf{Output:} Structured semantic rules $\mathcal{SR} = \{\mathcal{SR}_1, ..., \mathcal{SR}_n\}$
\begin{algorithmic}[1]
  \State $Text \gets \texttt{CleanDocument}(\mathcal{D})$
  \State $Paras \gets \texttt{SplitIntoParagraphs}(Text)$
  \State $\mathcal{R} \gets \emptyset$
  \ForAll{$para \in Paras$}
    \State $(p, m, c) \gets \texttt{LLM\_IdentifySR}(para, \mathcal{L})$
    \State $\mathcal{R} \gets \mathcal{R} \cup \{(p, m, c)\}$
  \EndFor
  \State $\mathcal{SR} \gets \emptyset$
    \ForAll{$\mathcal{R}_i \in \mathcal{R}$}
    \State $\mathcal{F} \gets \texttt{ExtractFieldPaths}(\mathcal{S}_i)$
    \State $(f,\; \mathcal{C},\; \mathcal{P})  \gets \texttt{LLM\_Complete}(\mathcal{R}_i,\ \mathcal{F})$
    \State $\mathcal{SR} \gets \mathcal{SR} \cup (p,\; m,\; f,\; \mathcal{C},\; \mathcal{P})$
    \EndFor
  \State \Return $\mathcal{SR}$
\end{algorithmic}
\end{algorithm}

\subsection{Semantic Rule Constructor}\label{sec: Rule Constructor}
To achieve semantic modeling, we introduce the structured rule constructor.
The overall procedure is summarized in Algorithm~\ref{alg:sr-constructor}, the process involves three main steps: data preprocessing, SR identification, and rule completion.
Specifically, in the preprocessing stage (lines 1–2), we first clean the raw document by removing non-essential content, and then segment the remaining text into paragraphs to preserve contextual coherence.
In the specification identification stage (lines 3–6), for each paragraph, a LLM is employed to extract a specification item $\mathcal{R}_i$.
All identified specification items are aggregated into the set $\mathcal{R}$.
In the rule completion stage (lines 8–12), based on each specification $\mathcal{R}_i$, we further derive its corresponding construction rule $\mathcal{C}$ and processing rule $\mathcal{P}$, thereby forming a structured rule $\mathcal{SR}_i$. Finally, all structured rules constitute the semantic rule set $\mathcal{SR}$, which provides a unified semantic foundation for subsequent analysis and verification.

\subsubsection{Data preprocessing}
To ensure accurate extraction of semantic rules, we preprocess each RFC document according to its identifier.
RFCs typically contain non-essential sections such as the table of contents, copyright notices, and references that are irrelevant to protocol semantics. We retain only the main body of the document, specifically the portion between the table of contents and the references, where most protocol definitions and operational details reside.
The remaining content is divided into coherent paragraphs to reduce LLM inference complexity and enable accurate rule extraction.
To preserve contextual relevance, each paragraph is prefixed with its hierarchical chapter titles, reflecting the original document structure.
Finally, each paragraph is represented as a key–value pair in the form {\texttt{chapter path}: \texttt{paragraph text}}.


\subsubsection{Specification Identification}
We design a specification identification module that leverages LLMs to analyze paragraph-level content and extract semantic knowledge from natural language descriptions.
Given a paragraph and a predefined message type set $\mathcal{L}$, the module identifies specification requirements $\mathcal{R} = (p, m, c)$, which characterize the normative behaviors of either the client or the server and serve as the foundation for subsequent semantic rule completion.
More details about the prompt design in this paper can be found in the Section~A of the Appendix.


\subsubsection{Rule Completion}
The rule completion module transforms each extracted specification $\mathcal{R}_i$ into a structured semantic rule $\mathcal{SR}_i = (p, m, f, \mathcal{C}, \mathcal{P})$.
Given the specification content $c$ and the corresponding seed message $\mathcal{S}_i$, the LLM leverages the message structure to identify the relevant field $f$ and derive two complementary components: the construction rule $\mathcal{C}$ and the processing rule $\mathcal{P}$.
This process converts textual specifications into explicit semantic constraints and behavioral descriptions, forming a consistent rule set $\mathcal{SR}$ for downstream mutation and verification tasks.


\subsection{Mutation Strategy Generator}\label{sec: Strategy Generator}
The mutation strategy generator transforms semantic rules into executable testing strategies that intentionally violate protocol constraints.
Given a semantic rule $\mathcal{SR}_i = (p, m, f, \mathcal{C}, \mathcal{P})$, the module employs a LLM to generate one or more mutation strategies $M = (p, m, f, d, e)$, where $d$ describes how to deviate from the construction rule $\mathcal{C}$, and $e$ represents the expected response inferred from the processing rule $\mathcal{P}$.
By converting compliant semantics into explicit violation cases, this process enables goal-directed fuzzing that systematically explores protocol boundaries and uncovers potential implementation flaws.
To ensure consistency in response evaluation, the LLM output is constrained to two high-level feedback categories: normal feedback (e.g., successful handshake) and error feedback (e.g., no response or abnormal termination).
More details can be found in the Appendix.

\subsection{Test Case Generator}\label{sec: Test Case Generator}
To accurately translate semantic intent into targeted test cases, we design a test case generator that applies structured mutation actions to real seed messages, ensuring that the generated cases remain valid while precisely targeting semantic boundary conditions.
As shown in Algorithm~\ref{alg:test-case-generator}, the generation process consists of two phases: action sequence generation and message mutation.

Specifically, for each mutation strategy $\mathcal{M}_i$ and its corresponding seed message $\mathcal{S}_i$, the module extracts the message structure $\mathcal{F}_i$, retaining only field paths (i.e., field names $k$) to provide contextual information (line 3).
Then, combining the mutation strategy and the message structure, the LLM generates an action sequence $\mathcal{A}_i$ (line 4), where each action is selected from a predefined set of atomic actions, \texttt{Add}, \texttt{Remove}, and \texttt{Update} (see Section~\ref{sec: asg}), to describe specific mutation behaviors.

In the message mutation stage (line 5), the execution engine applies the generated actions to the seed message to construct the test case $\mathcal{T}_i$. When an action contains incomplete parameters (e.g., missing length fields), the module automatically infers and fills them based on contextual information to ensure syntactic consistency.
All generated test cases are collected into the set $\mathcal{T}$, which serves as the input for subsequent response verification (line 6).

\begin{algorithm}[t]
\raggedright
\caption{Test Case Generation}
\label{alg:test-case-generator}
\textbf{Input:} Mutation strategies $\mathcal{M} = \{\mathcal{M}_1, ..., \mathcal{M}_n\}$, seed messages $\mathcal{S} = \{\mathcal{S}_1, ..., \mathcal{S}_n\}$ \\
\textbf{Output:} Test case set $\mathcal{T} = \{\mathcal{T}_1,\ \mathcal{T}_2,\ \dots,\ \mathcal{T}_n\}$
\begin{algorithmic}[1]
  \State $\mathcal{T} \gets \emptyset$
  \For{$i = 1$ \textbf{to} $n$}
    \State $\mathcal{F}_i \gets \texttt{ExtractFieldPaths}(\mathcal{S}_i)$
    \State $\mathcal{A}_i \gets \texttt{LLM\_GenerateActions}(\mathcal{M}_i,\ \mathcal{F}_i)$
    \State $\mathcal{T}_i \gets \texttt{MessageMutation}(\mathcal{S}_i,\ \mathcal{A}_i)$
    \State $\mathcal{T} \gets \mathcal{T} \cup \{\mathcal{T}_i\}$
  \EndFor
  \State \Return $\mathcal{T}$
\end{algorithmic}
\end{algorithm}

\subsubsection{Action Sequence Generation}\label{sec: asg}
Directly leveraging an LLM to modify hexadecimal message data often leads to invalid test cases.
This is particularly problematic in binary protocols such as TLS, where small changes in one field (e.g., an extension) may require cascading updates to other dependent fields (e.g., length indicators). 
To address this, we design the action sequence generation module, which transforms each mutation strategy into a high-level sequence of atomic actions.
Instead of directly manipulating raw bytes, the module specifies how to mutate a message in an abstract, structured manner.
The generated action sequence is then executed by a deterministic mutation engine that ensures structural integrity and automatically fills in undefined values.

Specifically, the module takes as input a mutation strategy $\mathcal{M}_i$ and the corresponding message structure $\mathcal{F}_i$ extracted from the seed message.
Each action belongs to one of the following types:
\begin{itemize}[leftmargin=*]
  \item \texttt{add(fields, position, value)}: Insert a field at the specified position; if the position is empty, the field is added to the end.
  \item \texttt{remove(fields)}: removes a field or subtree from the message based on its field path.
  \item \texttt{update(fields, new\_value)}: updates the value of a field; if it is empty, it is inferred during execution.
\end{itemize}
This design allows the LLM to focus on semantic-level mutation, while the execution engine maintains structural consistency, thereby bridging symbolic reasoning and byte-level manipulation.

\subsubsection{Message Mutation}
To ensure correctness and maintain syntactic validity, the execution of mutation actions is delegated to a deterministic message mutation engine. This component addresses the shortcomings of LLMs in byte-level accuracy and length-dependent computations.
The mutation engine takes as input a seed message and the corresponding action sequence.
For each action, it invokes predefined transformation functions to apply the requested modification to the nested message structure.
This structured representation preserves field hierarchy and allows precise updates.
When the action sequence contains undefined parameters, such as empty values in an \texttt{update} action, the mutation engine infers the missing content using the message context.
For instance, if an extension is inserted, the engine computes and updates the associated \textit{extension\_len} field accordingly.

\subsection{Response Verifier}\label{sec: Response Verifier}
To complete the fuzzing feedback loop, we design the response verifier to evaluate whether the implementation’s actual response aligns with the expected response $e$ defined in the semantic rule.
This closes the gap between mutation intent and response verification, enabling automated inconsistency detection.
During execution, the test case is sent by the role specified in the construction rule $\mathcal{C}$, and the response is observed from the corresponding receiver role.
The verifier monitors the interaction and records the implementation's actual response.
If any discrepancy is observed between the two, it is labeled as a potential vulnerability.

\section{Evaluation}
To evaluate the effectiveness of \tool{}, we aim to answer
 the following research questions:
\begin{enumerate}[label=\textbf{RQ\arabic*:}, widest=9, leftmargin=!, labelsep=.6em, labelindent=0pt]
    \item How effective is the LLM-based semantic modeling?
    \item How effective is \tool{} in detecting vulnerabilities compared to existing methods?
    \item How does each module in \tool{} contribute to the overall effectiveness of fuzzing?
    \item How does the choice of underlying LLM affect the overall fuzzing performance of \tool{}?
\end{enumerate}

\subsection{Experimental Setup}
\subsubsection{Baselines}\label{sec: baseline}
To evaluate the effectiveness of \tool{}, we selected four representative existing methods for comparison.
Among the grey-box methods, ChatAFL~\cite{CHATAFLPAPER} leverages the protocol syntax knowledge embedded in LLMs to guide fuzzing but is only applicable to plaintext protocols (e.g., HTTP).
On the other hand, Fuzztruction-Net~\cite{Nopeer} generates test cases by performing fault injection on one side of the communication.
However, this method relies on compile-time instrumentation of source code and is therefore applicable only to open-source protocol implementations.

Among the black-box methods, Hdiff~\cite{HdiffPaper} detects inconsistencies between protocol implementations through differential testing, but its effectiveness heavily depends on predefined templates.
In contrast, Bleem~\cite{BLEEM} infers states and guides testing by analyzing external network traffic, making it applicable to both open-source and closed-source protocol implementations.

In addition, to evaluate the contribution of each module in the framework, we designed two ablation variants.
First, to assess the role of the specification identification module in structured semantic modeling, we designed ${w/o SI}$, which directly extracts semantic rules from text.
Second, to evaluate the effect of action sequence generation, we designed ${w/o Action}$, which allows the large language model to directly modify seed messages.

\subsubsection{Benchmark dataset}\label{sec: baseline}
To objectively evaluate semantic artifacts, we constructed a benchmark dataset by systematically annotating the core RFC specifications of TLS 1.3, HTTP/1.1, IPv6, and DNS.
Eight cybersecurity master students independently cross-annotated each document, followed by expert adjudication to resolve disagreements and ensure semantic consistency.
The two-stage annotation process first extracted normative requirements from RFC text and then transformed them into structured semantic rules.
In total, the 28-hour effort yielded 1,721 high-quality semantic rules, comprising 624 from TLS 1.3, 540 from HTTP/1.1, 298 from IPv6, and 259 from DNS.


\begin{table}[t]
\centering
\small
\begin{threeparttable}
\caption{Details of our testing targets (*server version)}
\label{tab: targets}
\begin{tabular}{c|c c c c c}
\toprule
Target & Version & Protocol & RFC & Message Type \\ 
\midrule
Dns.exe      & *.143/*.206     & DNS      & 1035 & Binary \\ 
Tcpip.sys    & *.169/*.351     & IPv6     & 8200 & Binary \\ 
Schannel.dll & *.143/*.103     & TLS 1.3  & 8446 & Binary \\ 
Http.sys     & *.143/*.351     & HTTP/1.1 & 7230 & Text \\
Openssl      & 3.1.3           & TLS 1.3  & 8446 & Binary \\ 
LibreSSL     & 3.4.0           & TLS 1.3  & 8446 & Binary \\
Nginx        & 1.27.2          & HTTP/1.1 & 7230 & Text \\ 
\bottomrule
\end{tabular}
\end{threeparttable}
\vspace{-4mm}
\end{table}

\subsubsection{Experimental Environment}
All experiments were conducted on a high-performance workstation equipped with two 18-core 2.3 GHz Intel Xeon E5-2699 CPUs and 32 GB of memory.
Three virtualized environments were configured on this workstation to cover both closed-source and open-source network protocol stack implementations.
Specifically, the Windows test environment includes Windows Server 2022 (v20348) and Windows Server 2016 (v14393), used to evaluate the native network protocol components: \texttt{dns.exe} (DNS), \texttt{tcpip.sys} (IPv6), \texttt{schannel.dll} (TLS 1.3), and \texttt{http.sys} (HTTP/1.1), as listed in Table~\ref{tab: targets} (Rows 2–5).
In addition, we deployed three widely used open-source implementations on a Linux virtual machine running Ubuntu 20.04: \texttt{OpenSSL} and \texttt{LibreSSL} (for TLS 1.3), and \texttt{Nginx} (for HTTP/1.1), as shown in Table~\ref{tab: targets} (Rows 6–8).
All virtual machines share identical hardware configurations to ensure the fairness and comparability of experimental results.
All baselines were run under the same hardware configuration (CPU/RAM) and the same initial seeds, with 24 hours of exploration time.

\tool{} is based on the GPT-4o model~\cite{gpt4o}.
To control the randomness in LLM responses, we conducted multiple experiments under different parameter settings.
Based on empirical results, we selected the optimal configuration: the parameters top\_p and temperature were set to 0.1 and 0.5, respectively.
Since ChatAFL is implemented based on AFLNet~\cite{AFLNET}, we replaced it with NetAFL~\cite{WinAFL} when conducting experiments in the Windows environment.
To ensure a fair comparison, the base model of ChatAFL was also switched from GPT-3.5 to GPT-4o.
For Fuzztruction-Net~\cite{Nopeer}, we followed its default configuration.
Specifically, when testing OpenSSL and LibreSSL servers, the method employed their built-in client tools (e.g., s\_client) as the ``weird peers'' to generate test inputs, while testing the Nginx server used the client provided by the ngtcp2 project as the ``weird peer''.

\subsubsection{Evaluation Metrics}
We use precision, recall, and f1-score to evaluate the performance of rule extraction ($\mathcal{R}$) and transformation ($\mathcal{SR}$).
A generated rule is considered correct if it semantically matches an entry in the manually constructed benchmark, with strict consistency required for message and field names.
For downstream tasks, accuracy is used as the evaluation metric.
Due to the one-to-many nature of mutation strategy ($\mathcal{M}$) generation, it is impractical to build a complete benchmark; therefore, correctness is determined manually.
A strategy is regarded as correct if it effectively violates the corresponding rule and produces a verifiable response.
The accuracy of test cases ($\mathcal{T}$) is determined by their syntactic validity; a test case is deemed correct if it can be properly encoded and successfully sent.

\begin{table}[t]
\centering
\caption{Semantic modeling evaluation results}
\vspace{-2mm}
\begin{adjustbox}{width=\columnwidth} 
\begin{tabular}{cc|ccc|ccc|cc|cc}
\toprule
\multirow{2}{*}{RFC} &
  \multirow{2}{*}{Paras.} &
  \multicolumn{3}{c|}{$\mathcal{R}$} &
  \multicolumn{3}{c|}{$\mathcal{SR}$} &
  \multicolumn{2}{c|}{$\mathcal{M}$} &
  \multicolumn{2}{c}{$\mathcal{T}$} 
  \\ 
  \cmidrule{3-12} 
  & & P & R & F1 & P & R & F1 & Correct & Acc. & Correct & Acc. \\ 
  \midrule
1035   & 280 &  0.78  & 0.86  & 0.82  & 0.74  & 0.82 & 0.78  & 960 & 0.93  & 929   & 0.90  \\
8200  & 160 &  0.83  & 0.89  & 0.86  & 0.78  & 0.84 & 0.81  & 856  & 0.94 & 828    & 0.91 \\
8446  & 521 & 0.82 & 0.87 & 0.84  & 0.80  & 0.85 & 0.82  & 1,997   & 0.91  & 1,728  & 0.79 \\
7230 & 357 & 0.79 & 0.89 & 0.84  & 0.76  & 0.86 & 0.81  & 1,632   & 0.90 & 1,599  & 0.88 \\ 
\midrule
\multicolumn{2}{c|}{Average} & 0.80 & 0.88 & 0.84 & 0.77 & 0.84 & 0.80 & 1,361 & 0.92 & 1,271 & 0.87 \\
\bottomrule
\end{tabular}
\end{adjustbox}
\label{tab: rq1}
\vspace{-2mm}
\end{table}

\subsection{RQ1: Effectiveness of Semantic Modeling}
Table~\ref{tab: rq1} presents the detailed results of our semantic modeling process.
After processing all 1,318 RFC paragraphs, our framework demonstrates strong performance in both core stages: identifying specification requirements ($\mathcal{R}$) and transforming them into semantic rules ($\mathcal{SR}$).
The average F1-scores reach 0.84 and 0.80, respectively, indicating that \tool{} can accurately extract and structure protocol semantics from unstructured text.
The high-quality semantic rules provide a solid foundation for subsequent tasks.
Based on these rules, the downstream modules successfully generate 5,940 mutation strategies ($\mathcal{M}$) and test cases ($\mathcal{T}$), achieving average accuracies of 0.92 and 0.87, respectively.
This provides strong evidence of the effectiveness of our semantic modeling approach.

We also observe performance variations across protocols.
In particular, for TLS 1.3 (RFC 8446), the accuracy of test case generation drops to 0.79, which is notably lower than that of other protocols (all above 0.90).
This decrease mainly results from the structural complexity of TLS messages such as \textit{ClientHello}, whose deeply nested fields pose greater challenges to the reasoning capability of large language models, occasionally leading to mutation strategies that reference nonexistent fields.
Despite this variation, the overall performance of our framework remains strong, confirming the effectiveness of our approach in semantic rule modeling.

\find{\textbf{Answer to RQ1}: 
Our framework demonstrates high accuracy and F1-scores across all stages of semantic modeling.
The average F1-score remains no lower than 80\%, and the average accuracy exceeds 87\%, proving its effectiveness in semantic modeling and its strong capability to support semantics-aware fuzzing.
}

\begin{table*}[t]
\centering
\caption{Overview of the 10 confirmed vulnerabilities we found in different targets}
\vspace{-2mm}
\begin{adjustbox}{width=\textwidth}
\begin{threeparttable}
\begin{tabular}{c|ccccc}
\toprule
\textbf{Target} & \textbf{ID} & \textbf{Type} & \textbf{Description} & \textbf{Status} & \textbf{CVE} \\
\midrule
Dns.exe 
& 1 & Cache pollution & Insufficient validation of unsolicited records in DNS responses.  & Fixed & - \\
\midrule
\multirow{3}{*}{Tcpip.sys} 
& 2 & Integer overflow & Integer overflow from an unrecognized Option Type in Destination Options header. & Fixed  & CVE-2021-24074 \\
& 3 & Buffer overflow  & Buffer overflow from incorrect ESP header placement before the Fragment header.  & Fixed  & CVE-2022-34718 \\
& 4 & Integer overflow & Integer overflow from parsing a malformed Destination Options header.  & Fixed & - \\
\midrule
\multirow{2}{*}{Schannel.dll} 
& 5 & Uaf & Missing position check for the pre\_shared\_key extension in ClientHello. & Fixed & CVE-2023-28233 \\
& 6 & Uaf & Missing position check for the pre\_shared\_key extension in ClientHello. & Fixed & CVE-2023-28234 \\
\midrule
\multirow{4}{*}{Http.sys} 
& 7 & Tampering & Improper handling of extra whitespace between header field-name and colon. & Fixed & - \\
& 8 & Uaf & UAF caused by parsing random values in the Accept-Encoding header. & Fixed  & - \\
& 9 & Buffer overflow & Improper handling of HTTP Trailers. & Fixed  & - \\
& 10 & Integer overflow & Integer overflow from missing length validation on an excessively long ServiceName.
 & Fixed & - \\
\bottomrule
\end{tabular}
\begin{tablenotes}
\small
\item Note: \#2 and \#3 were discovered by an earlier version of SemFuzz (rule-based modeling), while the current version achieves the same results via LLM-based automated modeling.
\end{tablenotes}
\end{threeparttable}
\end{adjustbox}
\vspace{-2mm}
\label{tab: rq22}
\end{table*}

\begin{table}[t]
\centering
\caption{Unique vulnerabilities found by different methods}
\vspace{-2mm}
\label{tab: rq21}
\begin{adjustbox}{width=\columnwidth}
\begin{tabular}{c|cccccc}
\toprule
\textbf{Target} &
\textbf{ChatAFL} &
\makecell{\textbf{Fuzztruc}\\\textbf{tion-Net}} &
\textbf{Hdiff} &
\textbf{BLEEM} &
\textbf{\tool{}} &
\makecell{\textbf{Unique}\\\textbf{Vuls.}} \\
\midrule
Dns.exe       & - & - & - & \textcolor{gray}{0} & 1 & 1 \\
Tcpip.sys     & - & - & - & 1 & 3 & 3 \\
Schannel.dll  & - & - & - & \textcolor{gray}{0} & 2 & 2 \\
Http.sys      & 1 & - & 2 & 3 & 4 & 4 \\
OpenSSL       & - & \textcolor{gray}{0} & - & \textcolor{gray}{0} & \textcolor{gray}{0} & \textcolor{gray}{0} \\
LibreSSL      & - & \textcolor{gray}{0} & - & 1 & \textcolor{gray}{0} & 1 \\
Nginx         & \textcolor{gray}{0} & \textcolor{gray}{0} & - & \textcolor{gray}{0} & \textcolor{gray}{0} & \textcolor{gray}{0}\\
\midrule
\textbf{Sum}  & \textbf{1} & \textbf{0} & \textbf{2} & \textbf{5} & \textbf{10} & \textbf{11} \\
\bottomrule
\end{tabular}
\end{adjustbox}
\vspace{-2mm}
\end{table}

\subsection{RQ2: Comparison with Existing Methods}
\tool{} discovered a total of 16 potential vulnerabilities, among which 10 were confirmed by developers as real vulnerabilities, resulting in an accuracy of 62.5\%.
Table~\ref{tab: rq21} shows the number of real vulnerabilities detected by baseline methods: BLEEM (5), ChatAFL (1), Hdiff (2), and Fuzztruction-Net (0).
Among the 11 unique vulnerabilities collectively discovered by all tools, \tool{} covered 10 of them, demonstrating the best performance.

For gray-box methods, Fuzztruction-Net~\cite{Nopeer} relies on fault injection through source-code instrumentation, making it inapplicable to closed-source targets (e.g., \texttt{dns.exe}, \texttt{tcpip.sys}).
Moreover, due to its lack of semantic awareness, it fails to generate targeted test cases and did not discover any vulnerabilities even on open-source targets.
ChatAFL~\cite{CHATAFLPAPER}, although leveraging LLMs, is mainly designed for text-based protocols and cannot handle binary protocols such as DNS, IPv6, and TLS.
Furthermore, it lacks explicit semantic modeling, discovering only one vulnerability in \texttt{http.sys}.

For black-box methods, Hdiff~\cite{HdiffPaper} is specifically designed for the HTTP protocol and thus exhibits poor extensibility to other protocols.
Moreover, its heavy reliance on manually predefined templates limits its ability to explore boundary conditions, resulting in only two discovered vulnerabilities.
BLEEM~\cite{BLEEM} demonstrates relatively strong performance and detects 5 vulnerabilities.
Notably, it detected a unique vulnerability in LibreSSL.
This vulnerability involves memory corruption within the X.509 certificate parsing module, whose behavior is specified by independent standards such as RFC 5280.
Since our work focuses on modeling the semantics of the TLS 1.3 protocol based on RFC 8446, we did not detect this vulnerability.
Nevertheless, BLEEM lacks a deep understanding of RFC specifications, and its random mutation and state exploration strategies struggle to cover complex boundary conditions.
Moreover, its reliance on coarse-grained oracles (e.g., server crashes) further limits its performance, causing it to miss 5 vulnerabilities.

Table~\ref{tab: rq22} presents the detailed information of the 10 real vulnerabilities discovered by \tool{}.
These vulnerabilities have been confirmed by vendors, including 5 previously unknown vulnerabilities (\#1, \#2, \#3, \#5, \#6), for which 4 CVEs have been assigned.
In addition, 6 potential vulnerabilities were confirmed to be false positives: one caused by an incorrect action sequence generation, one triggered by an incorrect mutation strategy, two due to inconsistencies in the experimental configuration, and the remaining two were deemed non-vulnerabilities by developers because they were difficult to exploit.
More details can be found in the appendix.

\find{\textbf{Answer to RQ2}: 
\tool{} successfully discovered 10 vulnerabilities, including 5 previously unknown ones (with 4 CVEs).
In contrast, the best baseline detected only 5 vulnerabilities.
}

\begin{table}[t]
\centering
\caption{Effectiveness of each module in \tool{}}
\vspace{-2mm}
\begin{adjustbox}{width=\columnwidth}
\begin{tabular}{c|ccc|ccc|cc|cc}
\toprule
\multirow{2}{*}{Protocol} 
& \multicolumn{3}{c|}{\tool{}} 
& \multicolumn{3}{c|}{*${w/oSI}$} 
& \multicolumn{2}{c|}{\tool{}} 
& \multicolumn{2}{c}{*${w/oAction}$} \\
\cmidrule{2-7} \cmidrule{8-11}
& P & R & F1 & P & R & F1 & Correct & Acc. & Correct & Acc. \\
\midrule
DNS  & 0.74 & 0.82 & 0.78 & 0.69 & 0.79 & 0.74 & 929 & 0.90 & 539 & 0.52 \\
IPv6 & 0.78 & 0.84 & 0.81 & 0.75 & 0.78 & 0.77 & 828 & 0.91 & 198 & 0.22 \\
TLS 1.3  & 0.80 & 0.85 & 0.82 & 0.73 & 0.79 & 0.76 & 1,728 & 0.79 & 134 & 0.06 \\
HTTP/1.1 & 0.76 & 0.86 & 0.81 & 0.73 & 0.83 & 0.78 & 1,599 & 0.88 & 1,263 & 0.70 \\
\midrule
Average & 0.77 & 0.84 & 0.80 & 0.73 & 0.80 & 0.76 & 1,271 & 0.87 & 534 & 0.36 \\
\bottomrule
\end{tabular}
\end{adjustbox}
\label{tab: rq31}
\end{table}

\begin{table}[t]
\centering
\scriptsize
\caption{Ablation experiment results}
\vspace{-2mm}
\begin{adjustbox}{width=\columnwidth}
\begin{tabular}{c|ccc|ccc|ccc}
\toprule
\multirow{2}{*}{Target}  & \multicolumn{3}{c|}{\tool{}} & \multicolumn{3}{c|}{*${w/oSR}$} & \multicolumn{3}{c}{*${w/oAction}$} \\ \cmidrule{2-10} 
 & PVul. & RVul. & Acc. & PVul. & RVul. & Acc. & PVul. & RVul. & Acc. \\ \midrule
Dns.exe  & 2 & 1 & 0.50 & 2 & 1 & 0.50 & 0 & 0 & - \\
Tcpip.sys  & 4 & 3 & 0.75 & 3 & 2 & 0.67 & 0 & 0 & - \\
Schannel.dll  & 3 & 2 & 0.67 & 3 & 2 & 0.67 & 0 & 0 & - \\
Http.sys  & 4 & 4 & 1.00 & 3 & 3 & 1.00 & 4 & 2 & 0.5 \\ 
Openssl  & 3 & 0 & - & 3 & 0 & - & 0 & 0 & - \\ \midrule
Sum  & 16 & 10 & 0.63 & 14 & 8 & 0.57 & 4 & 2 & 0.5\\
\bottomrule
\end{tabular}
\end{adjustbox}
\label{tab: rq32}
\vspace{-2mm}
\end{table}

\subsection{RQ3: Ablation Study}
To evaluate the effectiveness of each core module in \tool{}, we conducted a series of ablation studies.
First, the specification identification module significantly improves the quality of semantic rule extraction from RFC documents.
As shown in Table~\ref{tab: rq31}, \tool{} achieves an average precision of 77\%, recall of 84\%, and F1-score of 80\% in the semantic rule extraction.
In contrast, the variant without this module (*$w/o SI$) attains an F1-score of 76\%.
This demonstrates that the module effectively filters irrelevant text and provides precise inputs for subsequent structured modeling, thereby enhancing the overall performance of semantic rule extraction.

Second, the action sequence generation module has a decisive impact on the quality of test case generation.
The results show that \tool{} generates an average of 1,271 correct test cases with an accuracy of 87\%.
When this module is disabled (*$w/oAction$), the system produces only 534 correct cases, and the accuracy drops sharply to 36\%. The performance gap becomes more pronounced for complex protocols such as IPv6 and TLS 1.3, where accuracy falls to 22\% and 6\%, respectively.
These results indicate that, without high-level action guidance, the LLM struggles to precisely locate specific fields and generate consistent byte-level mutations.

Finally, Table~\ref{tab: rq32} presents the contribution of each module to vulnerability detection.
\tool{} identifies 16 potential vulnerabilities, 10 of which are real vulnerabilities.
When the specification identification module is removed (*$w/oSI$), the number of confirmed vulnerabilities drops to 8, demonstrating the importance of fine-grained semantic modeling for vulnerability detection.
Moreover, disabling the action sequence generation (*$w/oAction$) severely degrades fuzzing effectiveness, yielding only 2 real vulnerabilities.
This result verifies the essential role of action sequences in translating mutation strategies into executable test cases.

\find{\textbf{Answer to RQ3}: 
The core modules of \tool{} enhance the overall performance.
The semantic rule constructor improves semantic modeling performance by approximately 5.3\% and detects 2 more vulnerabilities.
The test case generator further increases test case accuracy by 142\% and detects 8 more vulnerabilities.
}

\begin{table}[t]
\centering
\setlength{\tabcolsep}{2pt}
\caption{The impact of the LLM on overall performance}
\vspace{-2mm}
\begin{adjustbox}{width=\columnwidth}
\begin{tabular}{c|ccc|ccc|ccc|ccc}
\toprule
\multirow{2}{*}{Target}  & \multicolumn{3}{c|}{\tool{} (GPT-4o)} & \multicolumn{3}{c|}{*GPT-5} & \multicolumn{3}{c|}{*Gemini-1.5 Pro} & \multicolumn{3}{c}{*Gemini-2.5 Pro} \\ \cmidrule{2-13} 
 & PVul. & RVul. & Acc. & PVul. & RVul. & Acc. & PVul. & RVul. & Acc. & PVul. & RVul. & Acc. \\ \midrule
Dns.exe  & 2 & 1 & 0.50 & 1 & 1 & 1.00 & 1 & 0 & - & 1 & 1 & 1.00 \\
Tcpip.sys  & 4 & 3 & 0.75 & 4 & 3 & 0.75 & 5 & 3 & 0.60 & 4 & 3 & 0.75 \\
Schannel.dll  & 3 & 2 & 0.67 & 3 & 2 & 0.67 & 4 & 2 & 0.50 & 3 & 2 & 0.67\\
Http.sys  & 4 & 4 & 1.00 & 4 & 4 & 1.00 & 4 & 4 & 1.00 & 4 & 4 & 1.00\\ 
Openssl  & 3 & 0 & - & 3 & 0 & - & 4 & 0 & - & 3 & 0 & -\\ \midrule
Sum  & 16 & 10 & 0.63 & 15 & 10 & 0.67 & 18 & 9 & 0.50 & 15 & 10 & 0.67\\
\bottomrule
\end{tabular}
\end{adjustbox}
\label{tab: rq4}
\vspace{-2mm}
\end{table}

\subsection{RQ4: The Impact of Models on Fuzzing}
In this RQ, we replace the LLM GPT-4o with GPT-5~\cite{gpt5}, Gemini-1.5 Pro~\cite{Gemini1.5}, and Gemini-2.5 Pro~\cite{Gemini2.5}, and conduct experiments under the same settings.
As shown in Table~\ref{tab: rq4}, despite using different LLMs, all variants were able to discover at least 9 real vulnerabilities.
This indicates that the effectiveness of \tool{} primarily stems from its closed-loop and semantics-aware framework design, rather than relying on the unique capabilities of any specific model.
Moreover, even earlier models such as GPT-4o can cover most critical boundary conditions under the guidance of the framework.

The results also reveal performance differences caused by varying reasoning capabilities. For example, Gemini-1.5 Pro detected the largest number of potential vulnerabilities (18), but achieved the lowest precision (50\%), as it tends to adopt more aggressive mutation strategies. As the reasoning ability of the LLMs improves, the precision of vulnerability detection increases accordingly.
Gemini-2.5 Pro achieved the highest precision at 67\%.

\find{\textbf{Answer to RQ4}: 
The performance of \tool{} primarily originates from its closed-loop and semantics-aware design rather than from any specific LLM; stronger reasoning capabilities can further enhance its detection effectiveness.}
\section{Threat to Validity}

\textbf{Message-type Selection and Seed-traffic Collection}.
The effectiveness of the framework depends on two user-provided inputs: a list of protocol message types to test and corresponding seed traffic.
In this experiment we manually consulted the IANA registry to select 44 key message types across four protocols; this process took approximately 20 minutes.
More details can be found in the Section~B of the Appendix.
We then captured seed messages from standards-compliant client–server interactions to ensure precise scope control and syntactic correctness of the seed messages.

\textbf{Evaluation Metric Setup}.
We did not use traditional fuzzing metrics such as code coverage because \tool{} is designed for black-box scenarios and targets closed-source protocol implementations (e.g., tcpip.sys), for which instrumentation to obtain coverage is typically infeasible.
Instead, we focus on a more direct success criterion: the ability to discover real, verifiable vulnerabilities.
This vulnerability-centric evaluation both aligns with the ultimate goals of security testing and directly measures the framework’s ability to uncover deep semantic vulnerabilities.
\section{Related Work}

Recent research on protocol fuzzing can be broadly categorized into grey-box and black-box approaches.
Grey-box methods~\cite{TLSFuzzer, pham2020aflnet, SGFuzz, TCPFUZZ, CHATAFLPAPER, Nopeer, kao2025blueman, SnapFuzz, Stateafl} leverage instrumentation to infer protocol states and guide test generation. AFLNet~\cite{pham2020aflnet}, for instance, combines response-code inference with coverage feedback, while ChatAFL~\cite{CHATAFLPAPER} exploits LLM-derived syntax knowledge for text-based protocols.
Other techniques~\cite{Nopeer} inject faults into communication peers. However, their dependence on source-level instrumentation limits applicability to closed-source targets such as \textit{tcpip.sys}.

Black-box fuzzers~\cite{HdiffPaper, BLEEM, Snipuzz, Peach, chen2018rfc, 278336, 9286011} avoid instrumentation and thus suit closed-source settings. BLEEM~\cite{BLEEM} builds response-based state graphs, and Snipuzz~\cite{Snipuzz} adopts response-aware mutation.
However, lacking semantic guidance, they struggle to generate intent-driven inputs for boundary behaviors and rely on crashes as coarse oracles.
HDiff~\cite{HdiffPaper} uses RFC-guided differential testing to detect semantic divergences in HTTP, but relies on manual templates and offers limited semantic coverage.

Rule extraction from RFCs has been explored for security analysis and test automation~\cite{HDiff, pacheco2022automated, 281344, al2024hermes, ma2024one, LLMIF, chen2023ebugdec, Chen2022RIBDetectorAR, tian2019differential}.
For example, some studies~\cite{pacheco2022automated, al2024hermes} attempt to automatically extract finite state machines (FSMs) from specifications to support automated security analysis.
In addition, mGPTFuzz~\cite{ma2024one} and LLMIF~\cite{LLMIF} leverage LLMs to infer message structures of specific protocols (e.g., Zigbee) to guide testing.
However, these approaches lack fine-grained semantic modeling (such as field-level constraints), limiting their ability to uncover semantic vulnerabilities.
\section{Conclusion}
This paper proposes a semantic-aware black-box testing framework, \tool{}.
The framework leverages LLMs to extract structured semantic rules from RFC documents.
These rules not only guide the generation of intentional violation test cases but also identify precise semantic oracles, thereby addressing the limitations of existing approaches in semantic modeling, intent expression, and semantic vulnerability detection.
In the evaluation on seven widely used protocol implementations, \tool{}  discovered 16 potential vulnerabilities, 10 of which were confirmed, including 5 previously unknown vulnerabilities (with 4 assigned CVEs), demonstrating the effectiveness of our framework.

\section{Acknowledgments}
This work was partly supported by the national key research and development program of china (2023YFB3107100).

\clearpage

\bibliographystyle{ACM-Reference-Format}
\balance
\bibliography{ref}

\section*{Appendix}\label{sec: appendix}
In the appendix, we provide the supporting materials for this paper, including the prompt design, the message types involved in each protocol, the response mapping mechanism, and detailed information on the identified potential vulnerabilities.

\begin{table}[ht]
\scriptsize
\setlength{\tabcolsep}{3pt}
\centering
\caption{The chosen message type}
\begin{tabularx}{\columnwidth}{>{\centering\arraybackslash}m{0.15\columnwidth}|X|>{\centering\arraybackslash}m{0.1\columnwidth}}
\toprule
Protocol & \multicolumn{1}{c|}{Message Type} & Count\\
\midrule
DNS & \textcircled{1} DNS Query, \textcircled{2} DNS Response & 2\\
\midrule
\multirow{6}{*}{IPv6} & \textcircled{1} IPv6 header with Authentication Header, \textcircled{2} IPv6 header with ICMPv6, \textcircled{3} IPv6 header with TCP, \textcircled{4} IPv6 header with UDP, \textcircled{5} IPv6 header with Destination Options Header, \textcircled{6} IPv6 header with Encapsulating Security Payload Header, \textcircled{7} IPv6 header with Fragment Header, \textcircled{8} IPv6 header with Hop-by-hop Options Header, \textcircled{9} IPv6 header (with No Next header), \textcircled{10} IPv6 header with Routing Header &  \multirow{6}{*}{10} \\
\midrule
\multirow{15}{*}{TLS 1.3} & \textcircled{1} ClientHello with Application Layer Protocol Negotiation Extension, \textcircled{2} ClientHello with Application Settings Extension, \textcircled{3} ClientHello with Compress Certificate Extension, \textcircled{4} ClientHello with Ec Point Formats Extension, \textcircled{5} ClientHello with Encrypted Client Hello Extension, \textcircled{6} ClientHello with Extended Master Secret Extension, \textcircled{7} ClientHello with Key Share Extension, \textcircled{8} ClientHello with No Extension, \textcircled{9} ClientHello with Pre Shared Key Extension, \textcircled{10} ClientHello with Psk Key Exchange Modes Extension, \textcircled{11} ClientHello with Renegotiation Info Extension, \textcircled{12} ClientHello with Reserved Extension, \textcircled{13} ClientHello with Server Name Extension, \textcircled{14} ClientHello with Session Ticket Extension, \textcircled{15} ClientHello with Signature Algorithms Extension, \textcircled{16} ClientHello with Signed Certificate Timestamp Extension, \textcircled{17} ClientHello with Status Request Extension, \textcircled{18} ClientHello with Supported Groups Extension, \textcircled{19} ClientHello with Supported Versions Extension & \multirow{15}{*}{19} \\
\midrule
\multirow{9}{*}{HTTP/1.1} & \textcircled{1} http request with Host header, \textcircled{2} http request with request.line header, \textcircled{3} http request with User-Agent header, \textcircled{4} http request with Accept header, \textcircled{5} http request with Accept-Encoding header, \textcircled{6} http request with Authorization header, \textcircled{7} http request with http.cache\_control header, \textcircled{8} http request with Content-Type header, \textcircled{9} http request with Content-Length header, \textcircled{10} http request with Date header, \textcircled{11} http request with Last-Modified header, \textcircled{12} http request with Server header, \textcircled{13} http request with Location header & \multirow{9}{*}{13} \\
\bottomrule
\end{tabularx}
\label{tab:appendix1}
\vspace{-2mm}
\end{table}

\subsection*{A. Prompt Design for \tool{}} \label{sec: prompt design}
The prompt designs of our method are illustrated in Fig.~\ref{fig: p-1}, Fig.~\ref{fig: p-2}, Fig.~\ref{fig: p-3}, and Fig.~\ref{fig: p-4},  which correspond to specification identification, semantic rule completion, mutation strategy generation, and action sequence generation, respectively.

\begin{figure}[t]
\begin{lstlisting}[language=MyLang,basicstyle=\scriptsize]
@Persona: You are a protocol security expert. Your task is to identify protocol specification requirements from a given text.
@Instructions:
    @InputVariable:
        ${text}$
        ${message_type_list}$
    @Command Based on the provided message type list, identify all specific message types mentioned in the text.
    @Command For each identified message type, extract the specific content of the specification requirement from the text.
    @OutputVariable:
        ${specification_requirements}$
    @Format ...
    @Examples ...
\end{lstlisting}
\vspace{-2mm}
\caption{Prompt for specification identification}
\Description{}
\label{fig: p-1}
\end{figure}

\begin{figure}[t]
\begin{lstlisting}[language=MyLang, basicstyle=\scriptsize]
@Persona: You are a protocol security expert. Your goal is to translate the specification requirements into structured rules.
@Instructions:
    @InputVariable:
        ${specification_requirement}$
        ${message_structure}$
    @Command Identify the relevant message field associated with the specification requirement based on the message structure.
    @Command Analyze the specification requirement to identify both construction and processing rules.
    @Command Identify the roles in different types of rules.
    @Command Infer the missing rule in the specification requirement to complete the structured semantic rule.
    @OutputVariable:
        ${semantic_rule}$
    @Format ...
    @Examples ...
\end{lstlisting}
\vspace{-2mm}
\caption{Prompt for semantic rule completion}
\Description{}
\label{fig: p-2}
\end{figure}

\subsection*{B. Provided Message Types}
As shown in Table~\ref{tab:appendix1}, in our experiment, we provided key message types for each protocol.
These include 2 types of DNS messages, 10 types of IPv6 messages, 19 types of TLS 1.3 messages, and 13 types of HTTP/1.1 messages.

\subsection*{C. Response Mapping Mechanism} \label{sec: secC}
\tool{} defines a protocol-specific response mapping mechanism that maps actual and expected responses into ``normal'' or ``error'' categories.
The details are as follows:

For DNS, a response is classified as normal when the server returns data corresponding exactly to the queried domain. If the response includes unrelated domain records or indicates a query processing error, it is treated as an error case.
For IPv6, normal behavior is identified when the receiver replies with an ICMPv6 Echo Reply, indicating successful packet handling. Error cases include the return of ICMPv6 error messages, such as Time Exceeded or Parameter Problem, or the absence of a response.
For TLS 1.3, the exchange is considered normal when the server continues the handshake after receiving a ClientHello message, producing messages such as ServerHello.
Failure to respond or to proceed with the handshake is considered an error (e.g. Alert).
For HTTP, a normal case involves the server returning a success status code such as 200 OK, confirming proper request handling.
Error cases include responses with failure status codes, such as 400 or 500, or a complete lack of response.
This mapping mechanism enables \tool{} to consistently interpret implementation behavior across different protocols and detect semantic vulnerabilities.

\begin{figure}[ht]
\begin{lstlisting}[language=MyLang, basicstyle=\scriptsize]
@Persona: You are a protocol security expert.Your task is to generate rule violations based on the construction rules, and
then develop precise mutation strategies for those violations.
@Terminology:
    @Term Expected response List: "Normal feedback": "The server accepted the ClientHello message and proceeded with the handshake...", "Error feedback": "The server did not respond to the ClientHello message."...
@Instructions:
    @InputVariable:
        ${semantic_rule}$
    @Command Analyze the semantics of the construction rule to generate the direct violation of the construction rule.
    @Command Generate specific mutation strategies for the violation of the construction rule.
    @Command Based on the processing rules, select the appropriate expected behavior from the expected behavior list for each test strategies.
    @OutputVariable:
        ${mutation_strategies}$
    @Format ...
    @Examples ...
\end{lstlisting}
\vspace{-2mm}
\caption{Prompt for mutation strategy generator}
\Description{}
\label{fig: p-3}
\end{figure}

\begin{figure}[ht]
\begin{lstlisting}[language=MyLang, basicstyle=\scriptsize]
@Persona: You are a protocol security expert.Your task is to generate precise action sequences by analyzing the provided mutation strategies and message structures.
@Terminology:
    @Term Actions instructions:
        update():...; remove()...; add()...
@Instructions:
    @InputVariable:
        ${mutation_strategy}$
        ${message_structure}$
    @Command Analyze each mutation strategy to determine the necessary modification to the message.
    @Command For each necessary modification, choose the corresponding actions and complete its parameters.
    @Command Ensure that the sequence of actions maintains the integrity of the message structure...
    @OutputVariable:
        ${action_sequence}$
    @Format ...
    @Examples ...
\end{lstlisting}
\vspace{-2mm}
\caption{Prompt for action sequence generation}
\Description{}
\label{fig: p-4}
\end{figure}

\begin{table*}[htbp]
\centering
\caption{Detected potential vulnerability in Dns.exe, Tcpip.sys, and Schannel.dll}
\vspace{-2mm}
\begin{tabular}{cp{0.88\textwidth}}  
\toprule
\multicolumn{1}{c}{Target} & \multicolumn{1}{c}{Cause}\\
\midrule
\multirow{12}{*}{Dns.exe} & The vulnerability was exploited using a mutation strategy where an additional domain, not requested by the server, was inserted into the response message. This mutation resulted in a response containing not only the IP address corresponding to the server's requested domain but also the IP address linked to a not requested domain. The expected behavior was that the server would send an error feedback. However, during processing, the server only verified the presence of its requested domain within the response. If found, the server accepted the entire response and cached it, thereby unintentionally storing the IP address associated with the not requested domain. This oversight led to cache poisoning, as the server's cache was contaminated with invalid domain-IP mappings.\\
\cmidrule{2-2}
& The false positive was detected due to an atomic action sequence error, where the test case generator mistakenly modified ``www.example.com'' to ``example.com'' (instead of the correct ``WWW.EXAMPLE.COM''). It was expected that the server would reject this query and return an error. However, in reality, the server not only returned an IP address for subdomains of ``example.com'' but also for ``example.com'' itself. This inconsistency between the expected and actual behavior resulted in the false positive.\\
\midrule
\multirow{21}{*}{Tcpip.sys} & The vulnerability was identified when the mutation strategy added multiple AH headers to the original packet, ultimately resulting in an unusually long IPv6 message that was expected to be accepted by the server and server would send a normal feedback. However, when the server attempted to process the message, it crashed, causing a discrepancy between the next strategy expected which is a normal feedback and actual results and meaning to send an error feedback. Expert analysis confirmed that the crash was triggered by the inability of the IPv6 node to handle the oversized packet with a special payload field.\\
\cmidrule{2-2}
& The vulnerability was discovered when the mutation strategy placed the ESP header as part of the per-fragment, rather than as part of the fragment itself, leading to the generation of an IPv6 packet with the ESP header preceding the Fragment Header. This packet was expected to be rejected by the server and server would send an error feedback. However, during processing, the server crashed, causing a deviation from the next strategy expected outcome whose expected feedback is normal feedback. Expert analysis confirmed that the crash occurred due to a buffer overflow vulnerability triggered when the server attempted to reassemble the fragmented packet.\\
\cmidrule{2-2}
& The false positive was identified due to an error in the mutation strategy when generating the action sequence, which led to an incorrect output for the field corresponding to the modified message. This error resulted in the generation of a valid response message containing extra data, which was expected to be rejected by the server and server would send a normal feedback. However, when processing such messages, the server typically discarded the extraneous data and responded an error feedback, causing a discrepancy between the expected and actual outcomes and ultimately leading to the false positive.\\
\cmidrule{2-2}
& The vulnerability was identified due to the mutated message having an Option Type value exceeding 0x80. According to the RFC, the server should discard the packet and reply with an ICMP message with Code 2. However, the server did not immediately discard the packet. Instead, it continued to process the subsequent packet headers, which led to a buffer overflow vulnerability and caused the server to crash, resulting in a timeout in response.\\
\midrule
\multirow{13}{*}{Schannel.dll} & The vulnerability was identified due to a mutation strategy that placed the PSK extension before the supported\_versions field, resulting in a message where the PSK was not the final extension. The expected behavior was for the server to reject this message with an error feedback. However, upon processing the message, the server detected an anomaly and responded with a normal feedback, which deviated from the expected outcome. Upon expert investigation, it was found that the server performed an identity check on the PSK, which caused the constructed message to fail validation and trigger the normal feedback. However, the server did not check whether the PSK was the last extension, which should have led to a rejection with the error feedback if the PSK were not the final extension.\\
\cmidrule{2-2}
& The vulnerability was discovered in a manner similar to the previous one, but it applies to a scenario involving multiple servers with ticket-based verification, as opposed to the single-server binder validation scenario described earlier.\\
\cmidrule{2-2}
& The false positive was identified when the mutation strategy caused the client to send application data prior to the certificate exchange for authentication, with the expectation that the server would reject the packet and send an error feedback. However, the experimental setup involved a server that did not perform client authentication, allowing the server to continue communication with the client and successfully send a normal feedback.\\
\bottomrule
\end{tabular}
\label{tab: appendix2}
\end{table*}

\begin{table*}[htbp]
\centering
\caption{Detected potential vulnerability in Http.sys and Openssl}
\vspace{-2mm}
\begin{tabular}{cp{0.88\textwidth}}
\toprule
\multicolumn{1}{c}{Target} & \multicolumn{1}{c}{Cause}\\
\midrule
\multirow{17}{*}{Http.sys} & The vulnerability was discovered when the mutation strategy introduced an extra whitespace between the Content-Length header and the colon in the original packet before sending it to the server. The expectation was that the server would reject the malformed request and return an error response. However, instead of rejecting it, the server accepted the packet and responded with a normal response. After expert verification, this behavior was confirmed to be exploitable for HTTP Request Smuggling.\\
\cmidrule{2-2}
& The vulnerability was discovered when the mutation strategy injected a list of malformed or unknown content-coding tokens into the Accept-Encoding header. While the RFC requires the server to ignore unsupported encodings, the http.sys driver failed to properly validate the object during the parsing of these invalid tokens. This oversight triggered a UAF condition, causing the kernel to crash and preventing a normal HTTP response.\\
\cmidrule{2-2}
& The vulnerability was identified because the Trailer header fields generated by mutate contained multiple random values without actual meaning, and the request utilized chunked transfer encoding. According to the RFC, the server should correctly handle this packet and normally reply to the HTTP request. However, the reality is that the server crashed due to a buffer overflow vulnerability when processing these random trailer values, resulting in a timeout for the reply.\\
\cmidrule{2-2}
& The vulnerability is triggered because the length of the ServiceName generated by mutate in the request is very long. After accepting such a packet, the server should reply that the packet is normal. However, in reality, the server crashes due to an integer overflow vulnerability caused by incorrect length validation of the original ServiceName input, resulting in a timeout for the reply which is an error reply.\\
\midrule
\multirow{13}{*}{Openssl} & This false positive was triggered by the premature transmission of application data before the certificate exchange. While the fuzzer anticipated a rejection, the server accepted the data simply because client authentication was disabled in the experimental configuration.
\\
\cmidrule{2-2}
& 
Due to a mutation strategy that inserted the PSK extension before the supported\_versions field, the generated message had the PSK not as the final extension. The server was expected to reject this message and return an error feedback; however, it instead responded with normal feedback. 
However, since the vulnerability is difficult to exploit, the developer does not consider it a vulnerability.
\\
\cmidrule{2-2}
& When the mutation strategy added a new field — the 0xffff PSK key exchange mode — into the original packet and sent it to the server, the server was expected to reject the packet and return an error feedback. However, the server accepted the packet and responded with normal feedback. 
However, since the vulnerability is difficult to exploit, the developer does not consider it a vulnerability.
\\
\bottomrule
\end{tabular}
\label{tab: appendix3}
\end{table*}

\subsection*{D. Detected Potential Vulnerabilities}
As shown in Table~\ref{tab: appendix2} and Table~\ref{tab: appendix3}, we summarize the 16 potential vulnerabilities, among which 10 have been confirmed, including 5 previously unknown vulnerabilities.

\textbf{Dns.exe}.
Two potential vulnerabilities were identified.
One is a new cache pollution vulnerability caused by the server failing to verify the presence of extra domain names in responses.
The second, a false positive, resulted from the mutation strategy that produced a valid domain under the server’s authority, leading to normal processing.

\textbf{Tcpip.sys}.
Four potential vulnerabilities were identified.
Specifically, we identified a resource exhaustion vulnerability triggered by the processing of oversized packets with excessive \texttt{Authentication} Headers. Furthermore, two memory vulnerabilities were detected: a buffer overflow resulting from the incorrect placement of \texttt{ESP} headers during packet fragmentation, and an integer overflow occurring within the parsing logic of the \texttt{Destination Options} header.
The last one is a false positive, where the server discarded messages containing invalid extra data and continued normal operation.

\textbf{Http.sys}.
A total of four vulnerabilities were detected. The first involves improper handling of whitespace in the \texttt{Content-Length} header, enabling HTTP Request Smuggling. The next two include a UAF caused by parsing the \texttt{Accept-Encoding} field and a buffer overflow triggered by malformed HTTP \texttt{Trailers}. The final vulnerability stems from incorrect length validation of the \texttt{ServiceName}, causing an integer overflow.

\textbf{Schannel.dll}.
Three potential vulnerabilities were detected.
Two of them are previously unknown issues that have been assigned CVEs. Both are related to incorrect positioning or omission of the \texttt{PSK} extension in the \texttt{ClientHello} message, allowing malformed handshakes to succeed.
The remaining one was a false positive, where the client sent application data before authentication when client authentication was disabled, but it caused no harm.

\textbf{Openssl}.
Three potential vulnerabilities were detected.
The first involves the client sending application data prematurely, which was misidentified as a vulnerability because client authentication was disabled.
The second is an incorrect placement of the \texttt{PSK} extension, where the server failed to reject the abnormal handshake as required by the RFC.
The third is the absence of the \texttt{PSK} key exchange mode extension, causing the server to accept an invalid request.
Since the latter two potential vulnerabilities are difficult to exploit in practice, they are not considered real vulnerabilities.

\end{document}